\documentclass[12pt]{article}

 \newtheorem{theorem}{\bf Theorem}
 
  \newtheorem{remark}{\bf Remark}
\usepackage{amsfonts}
\usepackage{amsmath}
\usepackage{color}
\usepackage{natbib, url}

 \newcommand{\ba}{\begin{eqnarray}}
\newcommand{\ea}{\end{eqnarray}}
\newcommand{\bas}{\begin{eqnarray*}}
\newcommand{\eas}{\end{eqnarray*}}
\newcommand{\ben}{\begin{enumerate}}
\newcommand{\een}{\end{enumerate}}
\newcommand{\e}{ { \mathbb{E}}}
\newcommand{\var}{ {\mathbb{V}\rm ar }}

\newcommand{\pr}{ {\rm pr} }
\newcommand{\bit}{\begin{itemize}}
\newcommand{\eit}{\end{itemize}}

\def\T{{ \mathrm{\scriptscriptstyle \top} }}
\newcommand{\convergeto}{ {\overset{d}{\longrightarrow \; }}}

\begin{document}

\date{}
\title{A selective review on calibration information from similar studies based
on parametric likelihood or empirical likelihood}

\author{}
\maketitle

\vspace{-0.8in}

\begin{center}

{

{
By Jing Qin\\
National Institute of Allergy and Infectious Diseases, National Institutes of Health, USA
}

{
Yukun Liu\\
KLATASDS - MOE, School of Statistics, East China Normal University, China
}

{Pengfei Li\\
Department of Statistics and Actuarial Science, University of Waterloo, Waterloo, Ontario, Canada N2L 3G1}
}
\end{center}

%\hspace{10em}{\bf Jing Qin$^a$, Yukun Liu$^b$ and Pengfei Li$^c$}\\
%
%a=National Institute of Allergy and Infectious Diseases, National Institutes of Health, USA\\
%b=KLATASDS - MOE, School of Statistics, East China Normal University, China\\
%c=Department of Statistics and Actuarial Science, University of Waterloo, Canada\\

\begin{abstract}
\noindent
In multi-center clinical trials, due to various reasons,
the individual-level data are strictly restricted to be assessed publicly.
Instead, the summarized information is widely available from published results.
With the advance of computational technology, it has become very common
in data analyses to run on hundreds or thousands of machines simultaneous, with the data distributed across
those machines and no longer available in a single central location. How to effectively assemble  the summarized clinical data information
or information from each machine in parallel computation has become a challenging task for statisticians and computer scientists.
In this paper, we  selectively review some recently-developed statistical methods,
including   communication efficient distributed statistical inference, and
 renewal estimation and incremental inference,
which can be regarded as the latest development of calibration information methods
in the era of big data. Even though those methods were developed in different fields and in different statistical frameworks,
in principle, they are asymptotically equivalent to those well known methods developed in meta analysis.
Almost no or little information is lost compared with the case when full data are available.
As a general tool to integrate information, we also review the generalized method of moments and estimating equations approach by using
empirical likelihood method.

\noindent {\it Some key words}: Calibration information; Empirical likelihood;  Estimating equations; Generalized method of moments;  Meta analysis.
\end{abstract}

\section{Introduction}

Combining information from similar studies has been and
will be an extremely important strategy in statistical inference.
The most popular example of such methods is
meta analysis, which  pools the published results of multiple similar scientific studies
together to produce  an enhanced estimate without using the raw individual data from each study.
We refer to Borenstein et al. (2009) for a comprehensive introduction of meta analysis.
Due to various reasons such as privacy or capacity of computer storage
in massive data inference, only summarized data rather than the original individual data are available.
This poses a very challenging problem:
how to  conduct an efficient updated inference by
making full use of the summarized data?
In recent years,
 many methods of combining information
have been developed in economic studies,  machine learning, and
distributed statistical inferences.
The goal of this paper is to selectively review a few popular methods
that are able to integrate information in different disciplines.

Utilizing external summary data or auxiliary information to make a sharper inference  is an old and effective
method in survey sampling.   Due to restrictions such as cost effectiveness or
convenience, the variable of interest $Y$
may be available for a small portion of individuals. However, the explanatory variable $X$ associated with $Y$
may readily be available.  Cochran (1977) had a comprehensive discussion on the regression type estimators
by adapting the summarized information from $X$.
Chen and Qin (1993), Wu and Sitter (2001), and Chen et al. (2002)  used empirical likelihood (EL)  to incorporate such information in finite population.

With the advance of technology, many summarized statistical results are available in public domains. For example, many aggregated demographic and socioeconomic status data are given in the US census reports. The Surveillance, Epidemiology and End Results (SEER) program of the National Cancer Institute provides the population-based cancer survival statistics, such as covariate specific survival probabilities. Imbens and Lancaster (1994) combined Micro and Macro data
in economic studies through generalized method of moments (GMM).
Chaudhuri, Handcock and Rendall (2008) showed that the inclusion of the population level information can reduce bias and increase efficiency of the parameter estimates in a generalized linear model setup.
Wu and Thompson (2019) published an excellent monograph on combining auxiliary information in survey sampling.

In this paper, we will consider two situations.
First, the summarized  information was derived under the same statistical model.
Second, the summarized information was derived under similar
 but not exactly the same statistical models.
In general, combining information in the former case is easier.
The later case is more delicate since one has to take
  the heterogeneity among different studies into considerations.

The rest of this paper proceeds as follows.
In Section 2, we briefly review two simple and popular  meta methods
of  combining similar analysis results.
 As a general tool of synthesizing information from summarized information,
 we review Owen's (1988) EL method and
 Qin and Lawless (1994)'s over-identified parameter problem in Section 3.
 In particular, we present a new way of deriving the lower information bound
 for the over-identified parameter problem.
Section 4 discusses enhanced inference by utilizing auxiliary information.
Section 5 presents results on more flexible meta analysis where
the covariate information is collected differently even in similar
studies.  Calibrating information from previous studies is given in Section 6.
We discusses methods of using disease prevalence information
 for more efficient estimation in case and control studies in
Section 7.
The popular communication efficient distributed statistical inference in machine learning is discussed in Section 8.
Renewal estimation and incremental inference is briefly presented in Section 9.
Some discussions are provided in Section 10.

\section{Two simple combining information methods}

\subsection{Convex combination }
Suppose that
$\hat{\theta}_1$ and $\hat{\theta}_2$ are two asymptotically unbiased estimators
for $\theta$ from two independent studies,
and that they satisfy
\(
\sqrt{n}(\hat{\theta}_i-\theta)\sim N(0,\sigma_i^2),  i=1,  2
\).
The most straightforward way of  combining $\hat{\theta}_1$ and $\hat{\theta}_2$
is a convex combination,
\[
\hat{\theta}=\alpha \hat{\theta}_1+(1-\alpha)\hat{\theta}_2, \quad  0< \alpha <1.
\]
The asymptotic  variance of $\hat \theta$ is
$
\sigma^2=\alpha^2 \sigma_1^2+(1-\alpha)^2\sigma_2^2,
$
which takes its minimum at
\(
\alpha =\sigma_2^2/(\sigma_1^2+\sigma_2^2).
\)
This suggests   combining $\hat{\theta}_1$ and $\hat{\theta}_2$ by
\[
\hat{\theta}
=
 \frac{\sigma_2^2}{\sigma_1^2
 +\sigma_2^2}\hat{\theta}_1+\frac{\sigma_1^2}{\sigma_1^2+\sigma_2^2}\hat{\theta}_2
=
  \frac{\hat{\theta}_1/\sigma_1^2 + \hat{\theta}_2/\sigma_2^2}{  1/\sigma_1^2 + 1/\sigma_2^2 },
\]
an  inverse-variance  weighting estimator.
In general,  $\sigma_1^2$ and $\sigma_2^2$ are unknown, we may replace them by
their estimators $\hat\sigma_1^2$ and $\hat\sigma_2^2$ respectively, which leads to
\[
\hat{\theta}
=
 \frac{\hat{\sigma}_2^2}{\hat{\sigma}_1^2+\hat{\sigma}_2^2}\hat{\theta}_1+\frac{\hat{\sigma}_1^2}{\hat{\sigma}_1^2+
 \hat{\sigma}_2^2}\hat{\theta}_2
 = \frac{\hat{\theta}_1/\hat\sigma_1^2 + \hat{\theta}_2/\hat\sigma_2^2}{  1/\hat\sigma_1^2 + 1/\hat\sigma_2^2 }.
\]

As an alternative method, we may use the maximum likelihood method to argue that this is the best estimator.
We can treat $\hat{\theta}_i$   as an direct observation  from
\(
\hat{\theta}_i|\theta\sim N(\theta,\sigma_i^2)
\),  $i=1,  2$.
Then the  log-likelihood is (regarding $\sigma_1^2$ and $\sigma_2^2$ as known constants)
\[
 - (\hat{\theta}_1-\theta)^2/(2\sigma_1^2) -
 (\hat{\theta}_2-\theta)^2/(2\sigma_2^2).
\]
Maximizing this likelihood  with respect to $\theta$
or setting the score function to be zero,
we end up with the same inverse-variance  weighting estimator.

\subsection{Random effect meta analysis}

Dersimoni and Laird (1986) proposed a moment-based estimation method under
a random effect model for meta analysis.
Suppose
\bas
\hat{\theta}_i|\theta_i\sim N(\theta_i,w_i^{-1}), \quad
\theta_i\sim N(\theta,\tau^2), \quad i=1, 2, \ldots, K,
\eas
where $w_i^{-1}$s are treated as known.
Unconditionally  we have
$
\hat{\theta}_i\sim N(\theta,w_i^{-1}+\tau^2).
$
Consider the following inverse-variance weighting estimator for $\theta$,
\bas
\hat{\theta} =\frac{\sum_{i=1}^K \hat{\theta}_iw_i}{\sum_{i=1}^K w_i}
\eas
with variance
$
\var(\hat{\theta})= \sum_{i=1}^K w_i^2(w_i^{-1}+\tau^2)/(\sum_{i=1}^Kw_i)^2.
$
Define
\[
Q=\sum_{i=1}^Kw_i(\hat{\theta}_i-\hat{\theta})^2
=\sum_{i=1}^Kw_i(\hat{\theta}_i-\theta)^2 - (\hat{\theta}-\theta)^2 \sum_{i=1}^Kw_i.
\]
Easily we can check
\[
\e( Q ) =
(K-1)+\tau^2\left( \sum_{i=1}^Kw_i-\sum_{i=1}^Kw_i^2\Big/\sum_{j=1}^Kw_j \right),
\]
which implies that a natural estimator of  $\tau^2$ is
\[
\hat{\tau}^2=\frac{Q-(K-1)}{ \sum_{i=1}^Kw_i-\sum_{i=1}^Kw_i^2/\sum_{j=1}^Kw_j  }.
\]
For small sample sizes, there is no guarantee that this estimator
is non-negative;
one may replace it by
$
\max(\hat{\tau}^2,0).
$

Alternatively,  we may estimate $\tau$ using the likelihood approach.
The joint likelihood based on $\hat{\theta}_i$'s is
\[
\ell(\theta,\tau)=
 -\frac{1}{2}\sum_{i=1}^K   \frac{(\hat{\theta}_i-\theta)^2}{\tau^2+w_i^{-1}}
- \frac{1}{2}\sum_{i=1}^K  \log(\tau^2+w_i^{-1}).
\]
Maximizing $\ell$ with respect to $\theta$ and $ \tau^2$ gives
their maximum likelihood estimators (MLEs).

Lin and Zeng (2007) made comparisons on the relative efficiency of using summary statistics versus
individual-level data in meta-analysis. They found that in general there is no information loss by using the summarized information
compared with the inference based on original individual data if they are indeed available.

\section{Empirical likelihood and general estimating equations}

In this section we will briefly review Owen's (1988) EL and Qin and Lawless' (1994) estimating equations
approach since those methods have provided a general tool to assemble information from different sources.

The maximum likelihood method for regular parametric models has many optimality properties.
As a result, it is one of the most popular methods in statistical inference. However, model
mis-specification is a big concern since a misspecified model may lead to biased results.
When the underlying distribution is multinomial,
Hartley and Rao (1968) proposed a mean constrained estimator for
the population total in survey sampling problems.
To mimic the parametric likelihood but with robust properties,
Owen (1988, 1990) proposed the EL method,
which is a natural generalization of the multinomial likelihood when
the number of categories is the same as the sample size.
EL can be thought of as a bootstrap that does not
resample, and as a likelihood without parametric assumptions (Owen, 2001).

%To compute the profile likelihood of this general multinomial distribution
%which has atoms at data points, we can find
%the empirical likelihood method
%for constructing confidence regions  has sampling
%properties similar to bootstrap,  but the bootstrap uses re-sampling.

\subsection{Definition of empirical likelihood}

Suppose  $X_1,...,X_n$ are $n$ independent and identically distributed (iid) observations
from $X$ with the cumulative distribution distribution $F$. Without loss of generality, we assume there are no ties,
i.e.,  any two observations are unequal to each other.
Let
$
dF(X_i), \;i=1,2,...,n,
$
be the jumps of $F(x)$ at the observed data points.
The nonparametric likelihood is
\(
   L (F)=\prod_{i=1}^n p_i.
\)
It is clear that if any $p_i=0$, then $  L(F)=0$, and if $\sum_{i=1}^n p_i<1$, then
$  L(F) <   L(F_* )$,  where $F_*(x)= \sum_{i=1}^n p_iI(X_i\leq x)/\sum_{i=1}^n p_i$.
According to the likelihood principle (parameters with larger likelihoods are preferable),
one need only consider the distribution functions $F(x)$ with $p_i>0$ and $\sum_{i=1}^n p_i=1$.

If we maximize the log-likelihood
\ba
\label{log-el}
 \ell(F)=\sum_{i=1}^n \log p_i
\ea
 subject to the constraints
\ba
\label{el-constraint0}
\sum_{i=1}^n p_i=1, \quad  p_i\geq 0,
\ea
then we end up with
$
p_i=1/n, \; i=1,2,...,n.
$
Therefore the EL method estimates  $F$ by
\(
F_n(x)=\sum_{i=1}^n p_iI(X_i\leq x)=n^{-1}\sum_{i=1}^n I(X_i\leq x).
\)
This is the reason why the empirical distribution is called the nonparametric maximum likelihood estimator of $F(x)$.

Suppose we are interested in constructing a confidence interval for
$\mu=\e(X)=\int xdF(x)$, the mean of $X$. Since we have discretized $F$ at each of the observed data points, the
integral becomes $\mu=\sum_{i=1}^n p_i X_i$. Next we maximize the log nonparametric likelihood subject to an extra constraint
\ba
\label{constraint-mean}
\sum_{i=1}^n p_i(X_i-\mu)=0.
\ea
Maximizing the log-likelihood \eqref{log-el}
subject to the constraints \eqref{el-constraint0} and \eqref{constraint-mean},
the Lagrange multiplier method gives the profile log-likelihood of $\mu$,
%
%To implement this maximization, we can apply
%Assume that $\mu$ is an interior point of the
%convex hull spanned by the $n$ observations. Define
%\[
%h(\lambda_1,\lambda)=\sum_{i=1}^n \log p_i+\lambda_1(\sum_{i=1}^n p_i-1)-n\lambda\sum_{i=1}^n
%p_i(x_i-\mu).
%\]
%Taking derivative with respect to the $p_i$'s, we can easily show that
%\[
%p_i=\frac{1}{n}\frac{1}{1+\lambda (x_i-\mu)},
%\]
%where $\lambda$ satisfies the constraint equation
%\begin{equation}
%\sum_{i=1}^n \frac{x_i-\mu}{1+\lambda(x_i-\mu)}=0.\label{8.1.1}
%\end{equation}
%Since all $0<p_i<1$, in the univariate case, we can find that
%\[
%\frac{1-n^{-1}}{\mu-x_{(n)}}<\lambda<\frac{1-n^{-1}}{\mu-x_{(1)}},
%\]
%where $x_{(1)}$ and $x_{(n)}$ are the minimum and maximum observed values, respectively.
%In addition, the left hand side of the constraint equation  \eqref{8.1.1}  is monotone in $\lambda$. Hence the solution to  \eqref{8.1.1}  can be found  numerically by the bisection method. When $\mu$ is a vector, the constraint equation can be shown to be
%the derivative of a convex objective function. A modified Newton's method can be used
% to solve the equation. Once the value of $\lambda$ is obtained, the profile log-likelihood is
\begin{equation}
\ell_n(\mu)=-\sum_{i=1}^n \log\{ 1+\lambda^\T (X_i-\mu)\}-n\log n, \label{8.1.2}
\end{equation}
where $\lambda$ is the solution to
\(
\sum_{i=1}^n (X_i-\mu)/\{ 1+\lambda^\T (X_i-\mu) \}=0.
\)

We can treat $\ell_n(\mu)$  as a parametric likelihood  of $\mu$.
It is clear that based on this likelihood,
the maximum EL estimator of $\mu$ is
$\hat{\mu}=\bar{X}=n^{-1}\sum_{i=1}^n X_i$, which is exactly the sample mean.
Define the likelihood ratio function as
\[
R_n(\mu)=2\{\max_{\mu}\ell_n(\mu)-\ell_n(\mu)\}
=2\{\ell_n(\bar{X})-\ell_n(\mu)\}.
\]
Under the regularity conditions specified in Owen (1988, 1990),
as $n$ goes to infinity,
$
R_n(\mu_0)
$
converges in distribution to
the chi-square distribution with $p$ degrees of freedom,
where $p$ is the dimension of $\mu$ and $\mu_0$ is the true value of $\mu$.
%{\red Please note that  $p$ is assumed to be 1 in the above discussion. }

\subsection{General estimating equations}

The original empirical likelihood was mainly used
to make inference for linear functionals of the underlying population distribution
such as the population mean (Owen, 1988, 1990).
Qin and Lawless (1994) applied this method to  general estimating models,
which greatly  broadens its applications.
Specifically, suppose the  population of interest satisfies a
general estimating equation
\ba
\label{gee}
\e \{ g(X,\theta) \} = 0,
\ea
for a $r\times 1$ vector-valued function  $g$   and
some $\theta$, which  is a $p\times 1$ parameter to be estimated.
We assume $r\geq p$ as otherwise the true parameter value  of $\theta$ is undetermined.

For general estimating equations  with $r>p$
or  over-identified models,
Hansen (1982) proposed the celebrated
GMM,  which has become one of the most popular
methods in econometric literature.
In essence, the GMM minimizes
\[
\left\{ \sum_{i=1}^ng(X_i,\theta) \right\}^\T
\Sigma^{-1}
\left\{
\sum_{i=1}^ng(X_i,\theta) \right\}
\]
with respect to $\theta$, where $\Sigma$ is the variance matrix of the estimating equation $g(X,\theta)$.
If  $\Sigma$ is unknown, we may replace it by the sample variance
\(
\hat{\Sigma}=\frac{1}{n}\sum_{i=1}^ng(X_i,\tilde{\theta})g^\T(X_i,\tilde{\theta}),
\)
where $\tilde{\theta}$ is an initial consistent estimator of $\theta$.

Instead of GMM, Qin and Lawless (1994) used
the EL to make inferences for parameters
defined by a general estimating equation.
For discretized $F(x)$ satisfying \eqref{el-constraint0},
equation \eqref{gee} becomes
\ba
\label{el-gee}
\sum_{i=1}^n p_i g(X_i,\theta)=0.
\ea
Maximizing the log-likelihood \eqref{log-el} subject to
\eqref{el-constraint0} and \eqref{el-gee},
we have the profile log-likelihood of $\theta$  (up to a constant),
\bas
\ell_n (\theta)=-\sum_{i=1}^n \log\{ 1+\lambda^\T g(X_i,\theta)\},
\eas
where $\lambda$ is the Lagrange multiplier determined by
\(
\sum_{i=1}^n  g(X_i,\theta)/\{1+\lambda^\T g(X_i,\theta) \}=0.
\)
We then estimate $\theta$ by the maximizer   $\hat \theta = \arg\max_{\theta}\ell_n(\theta)$,
whose limiting distribution is established in the following theorem.
Hereafter we use $\nabla_{\theta}$ to denote the differentiation operator with respect to $\theta$.

\begin{theorem}[Qin and Lawless (1994)]
Denote $g=g(X,\theta_0)$ and   $\nabla_{\theta^\T} g=\nabla_{\theta^\T}g(X, \theta_0)$.
Suppose that  (1) $ \e( gg^\T) $ is positive definite,
(2) $\nabla_{\theta^\T} g(X,\theta) $ is continuous in a neighbourhood of $\theta_0$,
(3)  $\|\nabla_{\theta^\T }g(X, \theta) \|$ and
$\|g(X,\theta)\|^3$ can be bounded by some integrable function $G(X)$ in this neighbourhood,
and   (4)  $\e(\nabla_{\theta^\T}g )$ is of full rank. Then
as $n\rightarrow \infty$,
\(
\sqrt{n}(\hat{\theta}-\theta)\convergeto N(0,V),
\)
where $\convergeto $ stands for ``convergence in distribution" and
\ba
\label{V-el-gee}
V=\left \{ \e\left ( \nabla_{\theta} g^\T \right )  (\e gg^\T)^{-1}
\e\left ( \nabla_{\theta^\T} g \right )\right \}^{-1}.
\ea
\end{theorem}

\subsection{Information bound calculation}

How good can we estimate $\theta$ based on the over-identified parameter model
$\e\{ g(X,\theta) \} =0$?  Is  the maximum EL estimator optimal?
To answer these questions, we consider an ideal situation:
suppose the true underlying density $f(x,\theta)$ is known.
We can construct an enlarged parametric density model
\[
h(x,\eta, \theta)=\frac{\exp\{\eta^\T g(x,\theta)\}f(x,\theta)}{\int \exp\{\eta^\T g(t,\theta)\}f(t,\theta)dt},
\]
where  we have implicitly assumed
\(
\int \exp\{\eta^\T g(t,\theta)\}f(t,\theta)dt<\infty.
\)
Clearly $h(x,0, \theta)=f(x,\theta)$. In other words, $E_{\theta,\eta}\{g(X,\theta)\}=0$ if $(\theta,\eta)=(\theta_0,0)$,
where $\theta_0$ is the true value of $\theta$.
We shall show that even if the form of $f(x,\theta)$ is available,
the MLE of $\theta$ based on $h(x,\eta,\theta )$
has the same asymptotic variance  as
the maximum EL estimator.

With the parametric model $h$, we can estimate $\theta$ by maximizing
\(
L(\theta,\eta)=\prod_{i=1}^n h(X_i,\eta,\theta)
\)
with respect to $(\eta, \theta )$.
Denote the resulting  MLE by  $(\tilde{\eta}, \tilde{\theta})$.
We show in Section \ref{proof1} that as $n\rightarrow \infty$,
\ba
\label{V-enlarged-PL}
\sqrt{n}(\tilde{\theta}-\theta) \convergeto N(0, V),
\ea
where $V$ is defined in \eqref{V-el-gee}.
In general $f(x,\theta)$ is unknown, hence we expect  that the best estimator of $\theta$
 should have an asymptotic variance at least as large  as $V$.
Because the maximum EL estimator of $\theta$
of Qin and Lawless (1994) has the asymptotic variance $V$,
we conclude it achieves the lower information bound.

\begin{remark}
If $g(x,\theta)$ is  unbounded, we may construct a new density
\[
h(x,\theta,\eta)=\frac{\psi\{\eta^\T  g(x,\theta)\}f(x,\theta)}{\int \psi\{\eta^\T g(x,\theta)\}f(x,\theta)},
\]
where
$
\psi(x)=2(1+ e^{-2x})^{-1}
$ with
$
\psi(0)=\psi'(0)=1
$.
We may go through the same exercise to get the same conclusion.
\end{remark}

\begin{remark}
Back and Brown (1992) established a similar result by constructing an exponential family. In particular, they
defined
\(
h(x,\theta)=\exp\{\xi^\T(\theta) g(x,\theta_0)-a(\theta)\}f_0(x),
\)
where $f_0(x)=f(x,\theta_0)$ and $\xi(\theta)$ is determined implicity by the following conditions:
\bas
\xi(\theta_0)=0, \quad
 a(\theta_0)=0, \quad
\int \exp\{\xi^\T(\theta) g(x,\theta_0)-a(\theta)\}f_0(x)=1,
\eas
and
\bas
\int g(x,\theta)\exp\{\xi^\T(\theta) g(x,\theta_0)-a(\theta)\}f_0(x)dx=0.
\eas

In Back and Brown (1992) approach, $\xi(\theta)$ is determined implicitly by above  constraint equation, while in our new approach
$\eta$ is an independent parameter.
\end{remark}

\subsection{\label{proof1}A sketchy proof of \eqref{V-enlarged-PL}}
The log-likelihood based on the enlarged model $h(x,\eta,\theta)$ is
\bas
\ell =
\sum_{i=1}^n \{\eta^\T g(X_i,\theta)+\log f(X_i,\theta)\}-
n\log \left[\int \exp\{\eta^\T g(x ,\theta)\}f(x,\theta)dx\right].
\eas
The score functions evaluated at $(\theta,\eta)=(\theta_0,0)$ is
\bas
 \nabla_{\eta } \ell(\theta_0,0)
 &=& \sum_{i=1}^ng(X_i,\theta_0),\quad
 \nabla_{\theta } \ell(\theta_0,0) = 0,\quad
 \nabla_{\eta\eta^\T} \ell(\theta_0,0)
= -n \e ( g g^\T ),
\eas
and
\bas
 \nabla_{\eta\theta^\T } \ell(\theta_0,0)
&=& \sum_{i=1}^n\nabla_{\theta^\T} g(X_i,\theta_0)  -
n\e\{ \nabla_{\theta^\T} g(X,\theta_0) +g(X,\theta_0)(\nabla_{\theta^\T}\log f(X, \theta)) \}.
\eas

Under some mild assumptions such as that  $\int g(x,\theta)f(x,\theta)dx = 0$ holds for $\theta$
in a neighborhood of $\theta_0$,
differentiating   both of its sides  with respect to $\theta$ leads to
\bas
\e\{ \nabla_{\theta} g(X, \theta) \}
+ \e \{  \nabla_{\theta} g(X, \theta)
\nabla_{\theta} \log f(X, \theta) \}=0,
\eas
which means
$
 \nabla_{\eta\theta^\T } \ell(\theta_0,0)
= \sum_{i=1}^n \nabla_{\theta^\T} g(X_i,\theta_0).
$
Meanwhile if $f(x, \theta)$ satisfies some regularity conditions, then
\bas
\e[  \nabla_{\theta \theta^\T}  \log f(x,\theta_0)
+  \{\nabla_{\theta}  \log f(x,\theta_0)\}\{\nabla_{\theta}  \log f(x,\theta_0)\}^\T ]=0.
\eas
Therefore
\ba
\sqrt{n}\left (\begin{array}{c}
\tilde{\eta}-0\\
\tilde{\theta}-\theta_0
\end{array}
\right )
&=& \left (\begin{array}{cc}
-\e(gg^\T) & \e(\nabla_{\theta^\T} g ) \\
\e(\nabla_{\theta} g^\T ) & 0
\end{array} \right )^{-1}
\left (\begin{array}{c}
n^{-1/2}\sum_{i=1}^n g(X_i,\theta_0)\\
0
\end{array} \right )
+o_p(1).
\label{theta-eta}
\ea
This together with the fact that
$n^{-1/2}\sum_{i=1}^n g(X_i,\theta_0)
\convergeto N(0, \e(gg^\T))$ as $n$ goes to infinity
implies  \eqref{V-enlarged-PL}.

\subsection{Entropy family}

The enlarged parametric model satisfies
\[
\int h(x, \eta,\theta)g(x,\theta)dx=0,
\]
only if $\eta=0$. Naturally one may require  $\eta = \eta(\theta)$ to satisfy
\[
\int g(x,\theta)\exp\{\eta^\T g(x,\theta)\}f(x,\theta)dx=0.
\]
In the construction of the enlarged parametric model $h(x,\eta, \theta)$,
it  is often too restrictive to assume a known underlying parametric model $f(x, \theta)$.
We may replace the cumulative distribution function $F(x,\theta) = \int_{-\infty}^x f(t, \theta)dt$
by the empirical distribution $F_n(x)=n^{-1}\sum_{i=1}^n I(X_i\leq x)$.
In this situation,   $\eta = \eta(\theta)$ is the solution to
$
\sum_{i=1}^ng(x_i,\theta)\exp\{\eta^\T g(x_i,\theta)\}=0
$

Let $H(x, \eta, \theta) = \int_{-\infty}^x h(t, \eta, \theta)dt$.
For fixed parameter value $(\eta, \theta)$,
we have
\[
dH(X_i, \eta, \theta)
=  \exp\{ {\eta}^\T(\theta)g(X_i,\theta)\}/[ \sum_{j=1}^n \exp\{ {\eta}^\T(\theta)g(X_j,\theta)\} ],
\]
and the likelihood becomes
\[
\prod_{i=1}^ndH(X_i, \eta, \theta) = \prod_{i=1}^n \frac{\exp\{ {\eta}^\T(\theta)g(X_i,\theta)\}}{\sum_{j=1}^n \exp\{ {\eta}^\T(\theta)g(X_j,\theta)\}}.
\]
In fact this is equivalent to the EL  $\prod_{i=1}^n p_i$, where $p_i$'s
minimize the Kullback-Leibler divergence (up to a constant)
or  minus the exponential titling likelihood
$
 \sum_{i=1}^np_i\log (p_i)
$
subject to the constraint
$
\sum_{i=1}^n p_i=1
$,
$p_i\geq 0$,
and
$ \sum_{i=1}^n  p_i g(X_i,\theta)=0.
$
 See Susanne (2007) for more details.

\section{Enhance efficiency with auxiliary information }

In this section, we discuss methods of incorporating auxiliary information
to enhance estimation efficiency, which were also investigated by Qin (2000).
We assume  a parametric model $f(y|x,\beta)$
for the conditional density function of $Y$ given $X$,
and leave
the marginal distribution $G(x)$ of $X$ un-specified.
We wish to make inferences for $\beta$ when
  some auxiliary information is summarized
through an estimating equation
\bas
\e \{ \phi(X,\beta) \} = 0.
\eas
For example, if we know the mean  $\mu$ of $Y$,  then
we can construct  an estimating equation
$
\e (Y-\mu)=0.
$
We can take
\bas
\phi(X,\beta)=\int (y-\mu)f(y|X,\beta)dy=\int yf(y|X,\beta)dy-\mu.
\eas

Furthermore,  we assume that the response $Y$ may have missing values.
Let $D$ be the non-missingness  indicator, being
$1$ if $Y$ is available, and $0$ otherwise.
We assume a missing at random model
\[
P(D=1|Y=y,X=x) = P(D=1| X=x) =  \pi(x),
\]
where $\pi(x)$ depends only  on $x$.
Denote the observed data by $(d_i, d_iy_i, x_i)$ ($i=1, 2, \ldots, n$)
and  $p_i = dG(x_i) $.
The likelihood is
\bas
L&=&
\prod_{i=1}^{n} \{ \pi(x_i)f(y_i|x_i, \beta)dG(x_i) \}^{d_i}
[ \{1-\pi(x_i)\}dG(x_i) ]^{1-d_i} \\
&=&
\prod_{j=1}^{n} \{ \pi(x_j)\}^{d_j}\{1-\pi(x_j)\}^{1-d_j}
\cdot
\prod_{i=1}^{n} \{f(y_i|x_i, \beta)  \}^{d_i}
 \cdot p_i.
\eas
We can maximize this likelihood subject to the constraints
\bas
\sum_{i=1}^n p_i=1, \quad
 p_i\geq 0, \quad \sum_{i=1}^n p_i\phi(x_i,\beta)=0.
\eas
Since $\prod_{j=1}^{n} \{ \pi(x_j)\}^{d_j}\{1-\pi(x_j)\}^{1-d_j}$ is independent of $\beta$,
the profile hybrid  empirical log-likelihood (up to  a constant) is
\ba
\ell(\beta)=\sum_{i=1}^n [d_i\log f(y_i|x_i, \beta)-\log\{1+\lambda^\T\phi(x_i,\beta)\}],
\label{8.4.1}
\ea
where $\lambda$ is the Lagrange multiplier determined by
\ba
\sum_{i=1}^n \frac{\phi(x_i,\beta)}{1+\lambda^\T\phi(x_i,\beta)}=0. \label{8.4.c}
\ea
%The score function of $\beta$ is
%\bas
%\frac{\partial \ell}{\partial \beta}
%=\sum_{i=1}^n\left \{ d_i\frac{\partial \log f(y_i|x_i\beta)}{\partial \beta}
%-\frac{\lambda^\T\partial \phi(x_i,\beta)/\partial \beta}{1+\lambda^\T\phi(x_i,\beta)}
%\right \}.
%\eas
In the special case that missing data is completely at random, i.e., $\pi(x_i)$ is a constant,
 Qin (1992, 2000) established the following theorem.

\begin{theorem}
Let  $\beta_0$ be the true parameter value,
$\hat \beta$ be  the maximum hybrid  EL estimator, i.e.,
the maximizer of  \eqref{8.4.1}, and $\hat \lambda $ be the corresponding Lagrange multiplier.
Denote $\phi = \phi(X, \beta_0)$,
$ \nabla_{\beta} \phi   =  \nabla_{\beta} \phi(X, \beta_0), $
  and
\[
J  =  - \e\left\{d_i  \nabla_{\beta\beta^\T}  \log f(y_i|x_i, \beta_0)  \right\}
=\var\left\{d_i  \nabla_{\beta }\log f(y_i|x_i, \beta_0)  \right\}.
\]
Under some regularity conditions,
when $n$ goes to infinity, we have
\bas
\sqrt{n}((\hat{\beta}-\beta_0)^\T, \hat \lambda^\T)^\T \convergeto N(0,\Sigma),
\eas
where  $\Sigma = {\rm diag}( \Sigma_{11}, \Sigma_{22} )$ with
\ba
\label{Sigma11}
\Sigma_{11}
&=&
\{  J  + \e(\nabla_{\beta} \phi^\T)(\e\phi\phi^\T)^{-1} \e(\nabla_{\beta^\T } \phi )\}^{-1}, \\
\Sigma_{22}
&=&
\{\e(\nabla_{\beta} \phi^\T)J^{-1}\e(\nabla_{\beta^\T} \phi)
+\e(\phi\phi^\T)\}^{-1}.  \nonumber
\ea

\end{theorem}

\begin{remark}
Imbens and Lancaster (1994) studied the same problem  using GMM.
In particular, they directly combined the conditional score estimating equation
$\nabla_{\beta} \log f(y|x, \beta) $ and $\phi(x,\beta)$. Even though the first order large sample
results are the same, the hybrid EL based approach is more appealing since
it  respects the parametric conditional likelihood and
replaces only the marginal likelihood   by the EL.
Numerical comparison of the two methods was given in Qin (2000).
\end{remark}

\section{Combining summary information: A more flexible method in  meta analysis}

Meta analysis is a systematic way to combine published information.
The method has become very popular since little extra cost is needed.
The main restriction in meta analysis is that all studies must include
the same variables in the analyses. The only allowed difference is the
sample sizes. We have to discard some studies if they contain variables
different from others. Summarized information is available
from published results, such as census reports, national health studies, and so on.

Due to confidentiality or other reasons, we typically cannot gain access
to the original data except for the summarized reports.
Suppose we are interested in conducting a new study
that may contain some new variables of interest, which are not available
from the summarized information, for example, in genetic studies,
some new bio-markers and genes are newly discovered.
Below we discuss a more flexible method to combine published
information and individual study data for enhanced inference.
 Chatterjee et al. (2016) discussed a related  problem on the
 utilization of auxiliary information.  As Han and Lawless (2016)
pointed out, however, their methodology and theoretical results
were already developed by
Imbens and Lancaster (1994) and  Qin (2000) in the absence of selection bias sampling case.

We consider two cases. (I) Sample size for the summarized information
 is much larger than the sample size in the new study.
(II) Sample sizes from the two data sources are comparable.
In Case I,  we can treat the summarized information as known, i.e., the variation
  in the summarized data is negligible compared to the variation
  in the new study.
  In Case II, we have to take the variation in
  the summarized information into consideration since it is comparable to the variation in the new study.
We focus  on Case I in this section and  study Case II in  Section 6.

\subsection{Setup and solution}

Suppose the summarized results are based on the statistical analysis from response $Y$ and covariate variables $X$ (though the original data are not available), and in the new
study, in addition to $Y,X$, an extra covariate $Z$ is included. We are interested in fitting a parametric model
$f(y|x,z,\beta)$ for the conditional density function of $Y$ given $X$ and $Z$.
Let $(y_1^*,x_1^*),....,(y_N^*,x_N^*)$  be history data even though they are not available.
The published information can be summarized in two ways:
\bit
\item[(I)]
$
\bar{h}=N^{-1}\sum_{i=1}^Nh(y_i^*,x_i^*)
$
is known,
and
\item[(II)]
  $\gamma^*$ is the solution of an estimating equation
$
\sum_{i=1}^Nh(y_i^*,x_i^*,\gamma)=0,
$
where  the function $ h(y,x,\gamma)$ is known up to $\gamma$.
%was fitted with fitted value $\gamma^*$ as well as its standard error are available.
\eit

Let $(y_1,x_1,z_1),....,(y_n,x_n,z_n)$ be observed  data in the new study. The basic assumption is that
$(y_i,x_i), i=1,2...,n$ and $(y_i^*,x_i^*)$ have the same distribution.  To utilize the summarized information,
we can define estimating functions
 \[
 g =(g_1,g_3 ), \;\;g_1(y,x, z)=\nabla_{\beta} \log f(y|x,z,\beta), \;\; g_3(y,x)=h(y,x)-\bar{h},
 \]
  in Scenario (I), and
 \[
 g =(g_1 ,g_3 ), \;\;g_1(y,x, z)=\nabla_{\beta} \log f(y|x,z,\beta), \;\;
 g_3(y,x)=h(y,x,\gamma^*)
 \]
 in Scenario (II).
 We  consider only the situation that $n/N\rightarrow 0$. In other words, the variation in the auxiliary information is negligible.

 The empirical likelihood approach amounts to maximizing $\sum_{i=1}^n\log p_i$ subject to the constraint
 \[
 \sum_{i=1}^np_ig(y_i,x_i, z_i, \beta)=0, \;\; p_i\geq 0, \;\; \sum_{i=1}^np_i=1.
 \]
According to  Qin and Lawless (1994),
the asymptotic variance of the maximum EL estimator $\hat \beta$
based on estimating equations $g$ is
 \[
[ \e(\nabla_{\beta} g^\T ) \{ \e(gg^\T)\}^{-1} \e(\nabla_{\beta^\T} g) ]^{-1},
 \]
 where $ \nabla_{\beta} g = \partial g (y, x, z, \beta) /\partial \beta |_{\beta = \beta_0}$,
 $g= g (y, x, z, \beta_0) $,  and $\beta_0$ is the truth of $\beta$.
 Denote
 \[
 A=\e(gg^\T)=\left (\begin{array}{cc}
 A_{11} & A_{12}\\
 A_{12}^\T  & A_{22}
 \end{array} \right ), \;\; A_{22.1}=A_{22}-A_{12}^\T A_{11}^{-1}A_{12}.
 \]
 Equivalently the asymptotic variance can be written as
 \[
 [ \e(\nabla_{\beta} g_1^\T )A_{11}^{-1}\e(\nabla_{\beta^\T} g_1)
 + \e(\nabla_{\beta} g_1^\T)A_{11}^{-1}A_{12}A_{22.1}^{-1}A_{21}A_{11}^{-1} \e(\nabla_{\beta^\T} g_1)]^{-1},
 \]
 or
$
 (J+A_{12}A_{22.1}^{-1}A_{21})^{-1},
$
 where $A_{11}=J$ is the Fisher's information matrix.

 In the above approach the estimating equation $g_3=h(y,x)-\bar{h}$ does not involve the parameter
 $\beta$. This method may not be efficient. As an alternative approach, we define
 $  g_2(x,z,\beta) = \psi(x, z, \beta)$ with
 \[
\psi(x, z, \beta)= \e\{ h(Y,X)|X=x,Z=z \}-\bar{h} = \int h(y,x)f(y|x,z,\beta)dy-\bar{h}.
 \]
 Then
 \(
\e\{ g_2(x,z,\beta) \} =0.
 \)
 If we combine the empirical log-likelihood based on the estimating equation $g_2$ and the log-likelihood
 $\sum_{i=1}^n\log f(y_i|x_i,z_i,\beta)$ as in last section (See Equation \eqref{Sigma11}),
 then the asymptotic variance of the resulting MLE   $\hat \beta$ is given by
 \[
\{ J+\e(\nabla_{\beta} \psi^\T )(\e \psi\psi^\T)^{-1}\e(\nabla_{\beta^\T } \psi)\}^{-1}.
 \]

\subsection{A comparison}
Given the two pairs of estimation functions, $\{g_1, g_3\}$ and $\{g_1,g_2\}$,
we may wonder  combining which  pair   leads to a better estimator
if we directly compare their asymptotic variance formulae. Alternatively,
we may enquire whether we should combine all three constraints
$g=(g_1,g_2,g_3)$ together.
Write
$g_{12}= (g_1,g_2)$, $ a=\e\left\{ h^\T(y,x)\nabla_{\beta} \log f(y|x,z,\beta) \right\}$,
and
\bas
\e(gg^\T)=\left (\begin{array}{ccc}
J & 0 & a\\
0 &  \e(\psi\psi^\T) & \e(\psi\psi^\T)\\
a^\T & \e(\psi\psi^\T) & \e(hh^\T)
\end{array}
\right )=\left (\begin{array}{cc}
B_{11} & B_{12}\\
B_{12}^\T & B_{22}
\end{array} \right ),
\\
B_{11}=\left (\begin{array}{cc}
J & 0\\
0 & \e(\psi\psi^\T)
\end{array} \right ), \quad
 B_{12}=\left (\begin{array}{c}
a\\
\e(\psi\psi^\T)
\end{array}\right ).
\eas

Using the results  in Qin and Lawless (1994)  and
\[
\left (\begin{array}{cc}
B_{11} & B_{12}\\
B_{21} & B_{22}
\end{array} \right )^{-1}
=\left (\begin{array}{cc}
I & -B_{11}^{-1}B_{12}\\
0  & I
\end{array} \right )
\left (\begin{array}{cc}
B_{11}^{-1} & 0\\
0  & B_{22.1}^{-1}
\end{array} \right )
\left (\begin{array}{cc}
I & 0\\
-B_{21}B_{11}^{-1}  & I
\end{array} \right )
\]
with $B_{22.1}=B_{11}-B_{12}^\T B_{11}^{-1}B_{12}$,
we find that the asymptotic variance of $\hat{\beta}$ by combining
the  three estimating equations and
 $\sum_{i=1}^n\log f(y_i|x_i,z_i,\beta)$
is
\[
[J+ \e(\nabla_{\beta} \psi^\T ) \{ \e (\psi\psi^\T) \}^{-1} \e(\nabla_{\beta^\T } \psi)
+ \e(\nabla_{\beta} g_{21} )B_{11}^{-1}B_{12}B_{22.1}^{-1}B_{21}B_{11}^{-1}\e(\nabla_{\beta^\T} g_{12})]^{-1}.
\]

It can be shown that
\[
\e(\nabla_{\beta} g )=(-J, \e(\nabla_{\beta} \psi ), 0),
\quad
\e(\nabla_{\beta} g_{12})=(-J,a).
\]
Immediately, we have
\[
\e(\nabla_{\beta}g_{12} )B_{11}^{-1}B_{12}
=(-J,a)\left (\begin{array}{cc}
J^{-1} & 0\\
0 & \{ \e (\psi\psi^\T)\}^{-1}
\end{array} \right )
\left (\begin{array}{c}
a\\
\e(\psi\psi^\T)
\end{array} \right )=0,
\]
which implies that  the asymptotic variance
in the case of combining   $g_1,g_2$, and $g_3$
is the same as that in the case of combining only $g_1$ and $g_2$.
This indicates  that taking $g_3$ into account
leads to no efficiency gain
in the estimation of   $\beta$.

The method of combining $g_2$ and the
parametric likelihood $\prod_{i=1}^nf(y_i|x_i,z_i,\beta)$
is better than that of combining $g_1$,  $g_3$
and the parametric likelihood.
To see this, recall that the asymptotical variances for the MLEs of $\beta$ of the two methods
are
 \[
V_1 = \{ J+\e(\nabla_{\beta} \psi^\T )(\e \psi\psi^\T)^{-1}\e(\nabla_{\beta^\T} \psi )\}^{-1}.
 \]
 and
  \[
V_2 =  (J+A_{12}A_{22.1}^{-1}A_{21})^{-1}.
 \]
It suffices to show  that $V_2-V_1\geq 0$, namely $V_2-V_1$ is non-negative definite.

\subsection{Proof of $V_2-V_1\geq 0$}
For convenience, we assume that $\e(h) = 0$.
Because   $\e(\nabla_{\beta} \psi^\T )= A_{12}$ and $\psi = \e(h|X, Z)$,
it suffices to show
\ba
\label{inequality-A221}
A_{22.1} - \e (\psi\psi^\T)  = (A_{22}-A_{21}A_{11}^{-1}A_{12}) - \e [ \{ \e(h|X, Z)  \}^{\otimes 2} ]\geq 0.
\ea

Let $\e_* $ and $\var_* $ denote $\e(\cdot |X, Z)$ and $\var(\cdot |X, Z)$, respectively.
Because
\bas
\left (\begin{array}{cc}
A_{11} & A_{12}\\
A_{21} & A_{22}
\end{array} \right )
&=&
\e \left\{
\left (\begin{array}{cc}
g_{1} \\
h
\end{array} \right )^{\otimes 2} \right\}
=
\e\left\{ \var_*
\left (\begin{array}{cc}
g_{1} \\
h
\end{array} \right ) \right\}
+
\var\left\{ \e_*
\left (\begin{array}{cc}
g_{1} \\
h
\end{array} \right )   \right\}
\eas
and
$\e_*(g_1) = 0$,
it follows that
\bas
\left (\begin{array}{cc}
A_{11} & A_{12}\\
A_{21} & A_{22}
\end{array} \right )
&\geq&
\var\left\{ \e_*
\left (\begin{array}{cc}
g_{1} \\
h
\end{array} \right )   \right\}
=
\e
\left (\begin{array}{cc}
0  &  0 \\
0  &     \e_*(h)\e_*(h^\T)
\end{array} \right ).
\eas
Multiplying both sides by
$(-A_{21}A_{11}^{-1},   I  )$
from the left and by $(-A_{21}A_{11}^{-1},   I  )^\T$
from the right, we arrive at
\bas
 A_{22}-A_{21}A_{11}^{-1}A_{12}
\geq
\e\{ \e_*(h)\e_*(h^\T)\},
\eas
namely inequality \eqref{inequality-A221} holds,
which implies $V_2-V_1\geq 0$.

\section{Calibrate information from previous studies }

We consider calibrating information with parametric likelihood,
EL (Owen, 2001), and
GMM (Hansen, 1982).
When only summary information from previous studies is available,
these three well-known methods can be used to
calibrate such summary information
and to make inference about the unknown parameters of interest.
We may wonder whether there is efficiency loss in doing so
compared with the inferences based on the pooled data
as if they were all available.
Lin and Zeng (2014) found that  parametric-likelihood-based meta analysis
of  summarized information does not lose information
compared with the analysis based on individual data.
This is extremely important since individual data
may involve privacy issues, whereas summarized information does not.
We disclose that  not only parametric likelihood,
but also EL and GMM
own this nice property.

%Specifically, let
%$
%\{ Y_{ij}, \;  i=1,2,....,n_i\}
%$
%($j=1,2,...,  K$)
%be the original individual data from $K$ independent studies
%and suppose that  they are all independent and
%follow the parametric density funciton
%$
%f(y,\theta)
%$.
%If all of them are available, then the joint likelihood is
%$
%L(\theta)=\prod_{j=1}^K \prod_{i=1}^{n_j}f(y_{ij}, \theta).
%$
%Denote  $\tilde{\theta} = \arg\max L(\theta)$, the MLE of $\theta$.
%Similarly, the likelihood based on $j$-th group of study is
%$
%L_j(\theta)=\prod_{i=1}^{n_j}f(y_{ij},\theta).
%$
%Let $\hat{\theta}_j$ be the MLE based on $L_j(\theta)$.
%Lin and Zeng (2014) showed that $\hat{\theta}$,
%the optimal combination of $\hat{\theta}_j, j=1,2,...,K,$
% has the same asymptotic distribution as $\tilde{\theta}$.

\subsection{Efficiency comparison}

Suppose that  the data
$(Y_{ij}, X_{ij}) $ ($ j=1, 2, \ldots, n_i; i=1, 2, \ldots, K$) are
iid and satisfy one of the following assumptions:
\bit
\item[(I)]
\(
\pr(Y_{ij}=y|X_{ij}=x) =  f(y|x, \beta)
\),   or
\item[(II)]
\(
\e \{g(Y, X, \beta)\} = 0
\)
with $\beta_*$ being the true value of $\beta$.
\eit
Assume that data are available batch by batch,
and that $n_i/n = \rho_i\in(0,1)$ where $n=\sum_{i=1}^K n_i$.

For the $r$-th batch of data,
\bit
\item[(a)]
under assumption  (I),
 we define a parametric log-likelihood function
\bas
\ell_{r, \rm PL}(\beta)& =& \sum_{i=1}^{n_r} \log \{ f(Y_{ri}|X_{ri}, \beta) \};
\eas
\item[(b)]
under assumption  (II),
we define an empirical log-likelihood function
\bas
\ell_{r, \rm EL}(\beta)
& =&
\sup \left\{
\sum_{i=1}^{n_r} \log(n_r p_i):
p_i\geq 0, \;
\sum_{i=1}^{n_r} p_i =1, \
\sum_{i=1}^{n_r} p_i g(Y_{ri}, X_{ri}; \beta) = 0
\right\} \\
& =&
 - \sum_{i=1}^{n_r} \log\{1+\lambda_r^\T  g(Y_{ri}, X_{ri}; \beta)\},
\eas
 where
$\lambda_r$ satisfies
\(
 \sum_{i=1}^{n_r} \frac{ g(Y_{ri}, X_{ri}; \beta )}{1+\lambda_r^\T  g(Y_{ri}, X_{ri}; \beta )} = 0
\);
\item[(c)]
under assumption  (II),  we define the objective function of the GMM method
(GMM log-likelihood for short) as
 \[
\ell_{r, \rm GMM}(\beta)
 =
- \left\{ \sum_{i=1}^{n_r} g(Y_{ri}, X_{ri}; \beta) \right\}^\T
  \Omega^{-1}  \left\{ \sum_{i=1}^{n_r} g(Y_{ri}, X_{ri}; \theta) \right\},
 \]
 where $\Omega=\var\{ g(Y, X,\beta_*) \}$.
In practice, the $\beta_*$ in the expression of $\Omega$ is generally
replaced by a consistent estimator of $\beta$.
Using the truth $\beta_*$ of $\beta$ does not affect
the  theoretical analysis   in this section.

\eit

Let $\ell_r(\beta) = \ell_{r, \rm PL}(\beta)$, $\ell_{r, \rm EL}(\beta)$ or $\ell_{r, \rm GMM}(\beta)$.
Under certain regularity conditions, it can be verified that
for $\beta = \beta_* + O_p(n^{-1/2})$,
\ba
\label{approx-lr}
\ell_{r}(\beta)
& =&
U_r^{\T} \sqrt{n_r}(\beta- \beta_*)
-
\frac{n_r}{2}
 (\beta - \beta_*) ^\T V (\beta- \beta_*) + o_p(1).
\ea
In Case (a),
\bas
U_r =
 n_r^{-\frac{1}{2}} \sum_{i=1}^{n_r} \nabla_{\beta}\log \{ f(Y_{ri}|X_{ri}, \beta_*) \}, \quad
V  =
\var[\nabla_{\beta}\log \{ f(Y |X,  \beta_*) \}].
\eas
In Case (b)
\bas
U_r = n_r^{-\frac{1}{2}} \sum_{i=1}^{n_r}  g(Y_{ri}, X_{ri}; \beta_*), \quad
V =   A_{12} A_{22}^{-1}A_{21},
\eas
where
\bas
A  =
\left(\begin{array}{cc}
0  &  \e  \{ \nabla_{\beta}  g^\T(Y, X; \beta_*)\} \\
\e \{ \nabla_{\beta^\T}  g(Y, X; \beta_*)\} & \e \{  g(Y, X; \beta_*) g(Y, X; \beta_*)\}
\end{array}
\right)
\equiv
\left(\begin{array}{cc}
A_{11} & A_{12}  \\
A_{21} & A_{22}
\end{array}
\right).
\eas
In Case (c),
\bas
U_r &=& - \{\e \nabla_{\theta} g^\T(Y, X, \beta_*) \}\Omega^{-1}n_r^{-\frac{1}{2}}
 \sum_{i=1}^{n_r} g(Y_{ri}, X_{ri}, \theta_*), \\
V &=&    \{\e \nabla_{\theta} g^\T(Y, X, \beta_*) \}  \Omega^{-1} \{ \e \nabla_{\theta^\T} g(Y, X, \beta_*) \}.
\eas
Denote the MLE of $\beta$ based on the
$r$-th batch of data by
$
\hat \beta_r = \arg\max \ell_r(\beta).
$
The above approximation implies that
\[
\sqrt{n_r}(\hat \beta_r - \beta_*) =
V^{-1} U_r+ o_p(1) \convergeto N(\beta_*,  V^{-1}).
\]

When  the $K$-th batch of individual data  are available,
we are not accessible to the individual data of the previous $K-1$ batches any longer,
but  only have summarized information $(\hat \beta_j, \hat \Sigma_j), j=1, 2, \ldots, K-1$,
where $\hat \beta_j$ is the MLE based on the $j$th batch of data  and $\hat \Sigma_j = V^{-1}/n_j+o(n^{-1})$.
We can define an augmented log-likelihood
\bas
\ell_A(\beta)
&=&
\ell_K(\beta)
 - \frac{1}{2}
\sum_{j=1}^{K-1}
(\hat \beta_j - \beta)^\T  \hat \Sigma_j^{-1} (\hat \beta_j - \beta)
\eas
and the corresponding MLE
$
\hat \beta_{A} = \arg\max \ell_A(\beta).
$
For  $\beta = \beta_*+ O_p(n^{-1/2})$,
using the approximation in \eqref{approx-lr},
we have
\bas
\ell_A(\beta)
&=&
U_K^\T \sqrt{n_K} (\beta - \beta_*)
- \frac{n_K}{2}  (\beta - \beta_*) V  (\beta - \beta_*)  \\
&&
 - \frac{1}{2}
\sum_{j=1}^{K-1}
n_j (  \beta - \beta_*)^\T  V (  \beta - \beta_*)
+
\sum_{j=1}^{K-1}
n_j (\hat \beta_j - \beta_*)^\T  V (  \beta - \beta_*)
+ C+o_p(1) \\
&=&
n^{-1/2} \sum_{j=1}^{K }
\sqrt{n_j}  U_j^\T
\cdot \sqrt{ n } (  \beta - \beta_*)
 - \frac{n}{2}
 (  \beta - \beta_*)^\T  V (  \beta - \beta_*)
+  C+o_p(1),
\eas
where the constant $C$ is different from place to place.

For comparison,  based on the pooled data,
we define in Case (a) the parametric log-likelihood as
\bas
\ell_{\rm PL}(\beta)& =& \sum_{r=1}^K \sum_{i=1}^{n_r} \log \{ f(Y_{ri}|X_{ri}, \beta) \},
\eas
 define  in Case (b)  the empirical log-likelihood function as
\bas
\ell_{\rm EL}(\beta)
& =&
\sup \{
\sum_{r=1}^K  \sum_{i=1}^{n_r} \log(n  p_{ri}):
p_{ri}\geq 0, \;
\sum_{r=1}^K \sum_{i=1}^{n_r} p_{ri} =1, \
\sum_{r=1}^K \sum_{i=1}^{n_r} p_{ri} g(Y_{ri}, X_{ri}; \beta) = 0
\} \\
& =&
 -\sum_{r=1}^K \sum_{i=1}^{n_r} \log\{1+\lambda^\T  g(Y_{ri}, X_{ri}; \beta)\},
\eas
 where
$\lambda $ satisfies
\(
\sum_{r=1}^K \sum_{i=1}^{n_r} \frac{ g(Y_{ri}, X_{ri}; \beta )}{1+\lambda^\T  g(Y_{ri}, X_{ri}; \beta )} = 0
\),
and
in Case (c),   define the GMM log-likelihood as
 \[
\ell_{\rm GMM}(\beta)
 =
 -
 \left\{\sum_{r=1}^K \sum_{i=1}^{n_r} g(Y_{ri}, X_{ri}; \beta) \right\}^\T
  \Omega^{-1}  \left\{ \sum_{r=1}^K\sum_{i=1}^{n_r} g(Y_{ri}, X_{ri}; \theta) \right\}.
 \]

Let the  log-likelihood  based
on the pooled data be
$\ell_{\rm pool}(\beta) = \ell_{\rm PL}(\beta)$, $\ell_{\rm EL}(\beta)$,
and $\ell_{\rm GMM}(\beta)$ in Cases (a), (b), and (c), respectively.
It can be found that
\bas
\ell_{\rm pooled}(\beta)
=
n^{-1/2} \sum_{j=1}^{K }
\sqrt{n_j}  U_j^\T
\cdot \sqrt{ n } (  \beta - \beta_*)
 - \frac{n}{2}
 (  \beta - \beta_*)^\T  V (  \beta - \beta_*)
+  C  +o_p(1),
\eas
where $C'$ is a constant different from $C$.
Let $\hat \beta_{\rm pooled} = \arg\max \ell_{\rm pooled}(\beta)$.
By comparing $\ell_{\rm pooled}(\beta)$ and $\ell_{A}(\beta)$,
we arrive at
\bas
\ell_{\rm pooled}(\beta) = \ell_{A}(\beta)+  C  +o_p(1),
\eas
and
\bas
\sqrt{ n } ( \hat \beta_A - \beta_*)
&=& \sqrt{ n } ( \hat \beta_{\rm pooled} - \beta_*)+o_p(1) \\
&=& V^{-1}\cdot n^{-1/2} \sum_{j=1}^{K }
\sqrt{n_j}  U_j^\T +o_p(1) \\
&\convergeto&
N(0, V^{-1}).
\eas
This indicates that compared with the methods, including parametric likelihood, EL,
and GMM, based on all individual data,
the calibration method based on the last batch of individual data
and all summary results of the previous batches
has no efficiency loss.

\subsection{When nuisance parameters are present}
If for batch $i$, we assume that the data $(Y_{ij}, X_{ij}) $ ($j=1, 2, \ldots, n_i$)
satisfy either
\bas
\pr(Y_{ij}=y|X_{ij}=x) =  f(y|x, \beta, \gamma_i)
\eas
or
\bas
\e \{g(Y, X, \beta, \gamma_i)\} = 0,
\eas
where $\beta$ is common but $\gamma_i$ is a batch-specific parameter.
We define $\ell_r(\beta, \gamma_r)$ in the same way as $\ell_r(\beta)$.
Let $(\hat \beta_i, \hat \gamma_i)$ be the MLE
of $(\beta, \gamma_i)$ based on the $i$-th batch of data,
and assume that approximately
\bas
((\hat \beta_i - \beta )^\T, (\hat \gamma_i - \gamma_{i})^\T)^\T
\sim  N(0, \hat \Sigma_i)
\eas
with $\hat \Sigma_i = (\hat \Sigma_{i, rs}) _{1\leq r, s\leq 2}$.

We have two ways of combining information from previous studies.
If we use all the previous summary information, we can define
\bas
\ell_A^{(1)}(\beta, \gamma_1, \ldots,  \gamma_K)
&=&
\ell_K(\beta, \gamma_i)
 - \frac{1}{2}
\sum_{i=1}^{K-1}
((\hat \beta_i - \beta )^\T, (\hat \gamma_i - \gamma_{i})^\T)
\hat \Sigma_{i}^{-1} ((\hat \beta_i - \beta )^\T, (\hat \gamma_i - \gamma_{i})^\T)^\T.
\eas
As
$
\hat \beta_i |\hat \gamma_i
\sim N(\beta, \hat \Sigma_{i, 11\cdot 2} ),
$
where
$
 \hat \Sigma_{i, 11\cdot 2}
 =
  \hat \Sigma_{i, 11 }
  -
   \hat \Sigma_{i, 1  2}
    \hat \Sigma_{i, 2 2}^{-1}
     \hat \Sigma_{i, 21},
$
if using only this summary information, we can define
\bas
\ell_A^{(2)}(\beta, \gamma_K)
&=&
\ell_K(\beta, \gamma_K)
 - \frac{1}{2}
\sum_{i=1}^{K-1}
(\hat \beta_i - \beta)^\T  \hat \Sigma_{i, 11\cdot 2}^{-1} (\hat \beta_i - \beta).
\eas
Below we show that   the MLEs of $\beta$ based on these two likelihoods
are actually equal to each other. In other words,
there is no efficiency loss of estimating $\beta$ based on
$\ell_A^{(2)}(\beta, \gamma_K)$
instead of
$\ell_A^{(1)}(\beta, \gamma_1, \ldots,  \gamma_K)$.

To see this, it suffices to show
\ba
\label{likelihood-equal}
\sup_{\gamma_1, \ldots, \gamma_{K-1}}\ell_A^{(1)}(\beta, \gamma_1, \ldots,  \gamma_K)
=
\ell_A^{(2)}(\beta, \gamma_K).
\ea
Denote the inverse matrix of  $\Sigma_{i} $
by
$
\Sigma_{i}^{-1} = (\Sigma_{i}^{rs})_{1\leq r, s\leq 2},
$
where
\bas
 \Sigma_{i }^{11}
 &=&
   \Sigma_{i, 11\cdot 2 }^{-1}, \quad
 \Sigma_{i }^{21}
 =
- \Sigma_{i,22 }^{-1}\Sigma_{i, 21 }  \Sigma_{i, 11\cdot 2 }^{-1}, \quad
\Sigma_{i }^{12}
=
-  \Sigma_{i, 11\cdot 2 }^{-1}   \Sigma_{i, 12 } \Sigma_{i,22 }^{-1},\\
  \Sigma_{i }^{22}
 &=&
\Sigma_{i, 22 }^{-1} +
\Sigma_{i, 22 }^{-1}  \Sigma_{i, 21 }  \Sigma_{i, 11\cdot 2 }^{-1}
 \Sigma_{i, 12 } \Sigma_{i, 22 }^{-1}.
\eas
It can be seen that
\bas
\ell_A^{(1)}(\beta, \gamma_1, \ldots,  \gamma_K)
&=&
\ell_K(\beta, \gamma_K)
 - \frac{1}{2}
\sum_{i=1}^{K-1}
 (\hat \beta_i - \beta )^\T \Sigma_{i }^{11}(\hat \beta_i - \beta )   \\
 &&
+
\sum_{i=1}^{K-1}
 (\hat \beta_i - \beta )^\T \Sigma_{i }^{12}(\gamma_{i}-\hat \gamma_i )
 - \frac{1}{2}
\sum_{i=1}^{K-1}
(\gamma_{i}-\hat \gamma_i )^\T   \Sigma_{i }^{22}(\gamma_{i}-\hat \gamma_i ).
\eas
%Let $\nabla_{\gamma}$ denote the differentiation operator with respect to $\gamma$.
Setting $\partial \ell_A^{(1)}(\beta, \gamma_1, \ldots,  \gamma_K)/\partial \gamma_i = 0 $
($1\leq i\leq K-1$) gives
\bas
(\gamma_{i}-\hat \gamma_i ) =
(\Sigma_{i }^{22})^{-1}  \Sigma_{i }^{21} (\hat \beta_i - \beta ).
\eas
Putting this back in $\ell_A^{(1)}(\beta, \gamma_1, \ldots,  \gamma_K)$
gives
\bas
 \sup_{\gamma_1, \ldots, \gamma_{K-1}}\ell_A^{(1)}(\beta, \gamma_1, \ldots,  \gamma_K)
%&=&
%\ell_K(\beta, \gamma_K)
% - \frac{1}{2}
%\sum_{i=1}^{K-1}
% (\hat \beta_i - \beta )^\T \Sigma_{i }^{11}(\hat \beta_i - \beta )   \\
% &&
%+
%\sum_{i=1}^{K-1}
% (\hat \beta_i - \beta )^\T \Sigma_{i }^{12}(\Sigma_{i }^{22})^{-1}  \Sigma_{i }^{21} (\hat \beta_i - \beta ) \\
% &&
% - \frac{1}{2}
%\sum_{i=1}^{K-1}
%  (\hat \beta_i - \beta )^\T \Sigma_{i }^{12}
%(\Sigma_{i }^{22})^{-1}  \Sigma_{i }^{21} (\hat \beta_i - \beta ) \\
&=&
 \ell_K(\beta, \gamma_K)
 - \frac{1}{2}
\sum_{i=1}^{K-1}
 (\hat \beta_i - \beta )^\T
\{ \Sigma_{i }^{11} - \Sigma_{i }^{12}(\Sigma_{i }^{22})^{-1}  \Sigma_{i }^{21} \}(\hat \beta_i - \beta )+C  \\
&=&
\ell_K(\beta, \gamma_K)
 - \frac{1}{2}
\sum_{i=1}^{K-1}
 (\hat \beta_i - \beta )^\T
\Sigma_{i, 11\cdot 2}^{-1}(\hat \beta_i - \beta )+C,
\eas
where we have used the definition of $\Sigma_{i, 11\cdot 2}$ in the last equation.
We arrive at equation \eqref{likelihood-equal}
after comparing this with  the definition of  $\ell_A^{(2)}(\beta, \gamma_K)$.

\section{Use covariate specific disease prevalent information}

As discussed in the previous section, summarized statistics from previous studies can sometimes be utilized to enhance
the estimation efficiency in a current study. This is especially important in the big data era where many types of information can be found through internet.
More specifically, suppose the disease prevalence is known at various levels of a known risk factor $X$.
%This information may be available from  published results, or from meta-analysis or a large cohort study.
%We already showed how to use this type of information in Chapter 8 in   prospective studies.
In this section we combine this type of information in a case-control biased sampling setup.

\subsection{Induced estimating equations under case-control sampling}

The case-control sampling is one of
the most popular methods in cancer epidemiological studies. This is mainly due to
the fact
that it is the most convenient, economic and effective method. Especially in the study of rare diseases,  one has to collect large samples in order to get a reasonable number of cases by using prospective sampling, which may not be  practical.
Using the case-control sampling,
a pre-specified number of cases ($n_1$) and controls ($n_0$) are collected retrospectively from case
and control populations separately.
Typically this can be accomplished by sampling cases from hospitals,
and sampling controls from the general disease free population.

For a given risk factor $X$,  let $F_i(x) = \pr(X\leq x|D=i)$ for $i=0, 1$.
Given  $X$ in a range $(a,b]$, the disease prevalence is
\[
\pr(D=1|a<X\leq b)=\phi(a,b),
\]
where $\phi(a,b)$ is known. Using Bayes' formula we have
\[
\phi(a,b) = \frac{\pi \int_a^bdF_1(x)}{\pr(a<X\leq b)}, \quad
1-\phi(a,b) = \frac{(1-\pi)\int_a^bdF_0(x)}{\pr(a<X\leq b)}
\]
with $ \pi=\pr(D=1).$
It follows that
\[
\int_a^bdF_1(x)=\frac{1-\pi}{\pi}\frac{\phi(a,b)}{1-\phi(a,b)}\int_a^b dF_0(x),
\]
or
\[
\e_1\left [ I(a<X\leq b)\right ]=\frac{1-\pi}{\pi}\frac{\phi(a,b)}{1-\phi(a,b)}\e_0[I(a<X\leq b)],
\]
where $\e_0$ and $\e_1$ denote the expectation operators with respect to $F_0$ and $F_1$, respectively.

We assume that given covariates $X$ and $Y$, the underlying disease model is given by the
conventional logistic regression
\begin{equation}
\pr(D=1|x,y)=\frac{\exp(\alpha^*+x\beta+y\gamma+y x\xi)}{1+\exp(\alpha^*+x\beta+y\gamma+y x\xi)}. \label{14.4.0}
\end{equation}
Let $\alpha=\alpha^*-\eta$ with $\eta=\log \{(1-\pi)/\pi\}$.
It can be shown (See Qin, 2017) that this is  equivalent to the exponential tilting model
\[
f_1(x,y)=f(x,y|D=1)=\exp(\alpha+x\beta+y\gamma+y x\xi)f_0(x,y),
\]
where $f_0(x,y)=f(x,y|D=0)$.
As a consequence,
\bas
  \e_0\left\{  I(a<X\leq b)e^{
\eta+\alpha+\beta X+\gamma Y+\xi XY } \right\} =
 \frac{1-\pi}{\pi}\frac{\phi(a,b)}{1-\phi(a,b)}\e_0[I(a<X\leq b)],
\eas
or
\begin{equation}
\e_0\left\{ I(a<X\leq b)e^{\alpha+\beta X+\gamma Y+\xi XY }
-\frac{\phi(a,b)}{1-\phi(a,b)}I(a<X\leq b)\right\} =0. \label{14.4.1}
\end{equation}
Denote
\[
g_0(X,Y)=e^{ \eta+\alpha+\beta X_i+\gamma Y_i+\xi X_iY_i } -1,
\]
and the summarized auxiliary information equations as
\[
g_i(X,Y)=I(a_{i-1}<X\leq a_i)e^{\alpha+\beta X+\gamma Y+\xi XY }
-\frac{\phi(a_{i-1},a_i)}{1-\phi(a_{i-1},a_i)}I(a_{i-1}<X\leq a_i)
\]
with
$i=1,2,...,I$.
Then $\e_0 \{ g(X,Y) \}=0$, where
$
g(X,Y)=(g_0(X,Y),g_1(X,Y),....,g_I(X,Y))^\T.
$

\subsection{Empirical likelihood approach }

The log-likelihood is
\ba
\label{el-case-control}
\ell=\sum_{i=1}^n d_i(\eta+\alpha+\beta x_i+\gamma y_i+\xi x_iy_i) +\sum_{i=1}^n \log (p_i),
\ea
where $p_i=dF_0(x_i), i=1,2,...,n$, and the constraints are
\bas
p_i\geq 0, \quad  \sum_{i=1}^n p_i=1,\quad
\sum_{i=1}^n p_i g(x_i,y_i)=0.
\eas
The profile log-likelihood is
\[
\ell=\sum_{i=1}^n d_i(\eta+\alpha+\beta x_i+\gamma y_i+\xi x_iy_i)
-\sum_{i=1}^n \log\{ 1+\lambda^\T g(x_i,y_i)\},
\]
where the Lagrange multiplier $\lambda$ is determined by
\[
\sum_{i=1}^n \frac{g(x_i,y_i)}{1+\lambda^\T g(x_i,y_i)}
=0.
\]  Finally, the underlying parameters can be obtained by maximizing $\ell$.

If the overall disease prevalence probability $\pi=\pr(D=1)$ is known, then
$\eta=\log \{(1-\pi)/\pi\}$ is known.
On the other hand if it is unknown but $I\geq 1$, then $\pi$ is identifiable. If $I>1$,
then we have an over-identified equation problem.
This can be treated as a generalization of the empirical likelihood method
for estimating functions (Qin and Lawless, 1994) to biased sampling problems.
Qin et al. (2015) considered the case that $\eta$ is unknown and $I\geq 1$.

Let
$
\omega=(\eta,\alpha,\beta,\gamma,\xi,\lambda).
$
Since the first estimating function $g_0$ corrects biased sampling in a case-control study, the remaining estimating
functions $g_1,...,,g_I$ are used for improving efficiency.
When $n$ goes to infinity,  it can be shown that  the limit  of $\lambda$ is
a $(I+1)$-dimensional vector with the first component being $\lim\limits_{n\rightarrow\infty} (n_1/n)$
and  the rest all being zero.
 Qin et al. (2015) disclosed that  if $\rho=n_1/n_0$ remains constant as $n\rightarrow \infty$ and $\rho\in (0,1)$,
 then under suitable regularity conditions $\sqrt{n}(\hat{\omega}-\omega_0)$ is asymptotically normally distributed
with mean zero.
Moreover, the estimation of the logistic regression parameters $(\beta,\gamma,\xi)$
improves as the number $I$ of estimating functions increases.
This means that   a richer set of auxiliary information leads to better estimators.
In practice, however, this must be balanced with the numerical difficulty of
solving a larger number of equations.

It is interesting to note that, auxiliary information is primarily informative for estimating $\beta$ and $\xi$, but not for estimating $\gamma$. This can be observed through the following equations
\bas
&& \int I(a<x<b)\exp(\alpha+\beta x+\gamma y+\xi xy)dF_0(x,y)\\
&& \hspace{5cm} =
\int I(a<x<b)\exp(\alpha+\beta x+s+\xi xs/\gamma)dF_0(x,s/\gamma).
\eas
Since the underlying distribution $F_0(x,y)$ is unspecified,
we can treat $F_0(x,s/\gamma)$ as a new underlying distribution $F_0^*(x,s)$.
With $F_0^*$  profiled out, the auxiliary information equation does not involve $\gamma$
if $\xi=0$. Hence, even if $\xi\neq 0$, the information for $\gamma$ is minimal since $\gamma$ and $\xi$ cannot be separated.

\subsection{Generalizations}

In Qin et al. (2015)'s simulation studies, it looks like the
maximum reduction of variance occurs for the estimation of the
 coefficient of $X$. If the auxiliary information
\[
\pr(D=1|b_{j-1}<Y\leq b_j)=\psi_j, \;\; j=1,2,...,J
\]
is also available, naturally we can combine them through estimating equations
\bas
g_i(X,Y)
&=& I(a_{i-1}<X\leq a_i)e^{\alpha+\beta X+\gamma Y+\xi XY) }
-\frac{\phi(a_{i-1},a_i)}{1-\phi(a_{i-1},a_i)}I(a_{i-1}<X\leq a_i), \\
h_j(X,Y)
&=&
I(b_{j-1}<Y\leq b_j)e^{\alpha+\beta X+\gamma Y+\xi XY }
-\frac{\psi(b_{j-1},b_j)}{1-\psi(b_{j-1},b_j)}I(b_{j-1}<Y\leq b_j).
\eas
It would be more informative if the auxiliary information
$\pr(D=1|a<X<b,c<Y<d)$
is available.

\subsection{More on the use of auxiliary information }

Under a logistic regression model,
the case and control densities are linked by the exponential tilting model
\ba
\label{logistic}
\pr(x,y|D=1)=\pr(x,y|D=0)\exp(\alpha+x\beta+y\gamma+\xi xy).
\ea
Suppose that for the general population
$
\e(X)=\mu_1
$,
$ \e(Y)=\mu_2
$ and
$ \e(XY)=\mu_3,
$
are all known,
and $\pi=\pr(D=1)$ is known or can be estimated using external data.
Under the exponential tilting  model \eqref{logistic}, the density $f(x,y)$ in the general
population and the density $\pr(x,y|D=0)$ in the control population are linked by
\begin{eqnarray*}
\pr(x,y)
&=&  \{\pi e^{\alpha+x\beta+y\gamma+\xi xy}+(1-\pi)\}\pr(x,y|D=0).
\end{eqnarray*}
As a consequence
\bas
\e(X)
&=& \e_0[X\{\pi e^{ \alpha+X\beta+Y\gamma+\xi XY}+(1-\pi)\}]=\mu_1,
\eas
where $\e_0$ is an expectation with respect to $\pr(x,y|D=0)$.
Let
$
h(x,y)=(x-\mu_1,y-\mu_2,xy-\mu_3)
$
with known $\mu_1,\mu_2$ and $\mu_3$.
The  log-likelihood under case-control data is still
\eqref{el-case-control},
where $p_i$'s satisfy the following constraints
\bas
&& \sum_{i=1}^np_i=1, \;\; p_i\geq 0,
\quad
\sum_{i=1}^np_i e^{ \alpha+x_i\beta+y_i\gamma+x_iy_i\xi} = 1,
\\
&&
\sum_{i=1}^n p_i h(x_i,y_i)\{\pi  e^{ \alpha+x_i\beta+y_i\gamma+x_iy_i\xi} +(1-\pi)\}=0.
\eas

More generally, any information in the general population such as
\(
\e[\psi(Y,X)]=0
\)
can be converted to an equation for  the control population,
\bas
\e_0[\{\pi e^{ \alpha+X\beta+Y\gamma+\xi XY} +(1-\pi)\}\psi(Y,X)]
=0.
\eas
Therefore the results developed in Qin et al. (2015) can be applied too.
Chatterjee et al. (2016)'s results for case-control data can be considered as a special case
of Qin et al. (2015).

\section{Communication efficient distributed   inference}

In the  era of big data,   it is commonplace
for data analyses to run on hundreds or thousands of machines, with the data distributed across
those machines and no longer available in a single central location.
Recently the parallel and distributed inference has become popular
in statistical literature both for frequentist settings and Bayesian settings.
In essence the data-parallel procedures are to break the  overall dataset into subsets
that are processed independently. To the extent that communication-avoiding procedures have been
discussed explicitly, the focus has been on one-shot or embarrassingly-parallel approaches that
only use one round of communication in which estimators or posterior samples are first obtained in parallel on
local machines, are then communicated to a center node, and finally  are
combined to form a global estimator or
approximation to the posterior distribution
(Zhang et al., 2013, Lee et al., 2017, Wang and Dunson,
2015, Neiswanger et al., 2015).
In the frequentist setting,
most one-shot approaches rely on averaging
(Zhang et al., 2013), where the global estimator is
the average of the local estimators. Lee et al. (2017)
extends this idea to high-dimensional sparse linear regression by
combining local debiased Lasso estimates (van de Geer et al., 2014).
Recent work by Duchi et al. (2015) shows that under certain conditions,
these averaging estimators can attain the information-theoretic complexity lower
bound for linear regression, and at least $O(dk)$ bits
must be communicated in order to attain the minimax rate of parameter estimation,
where $d$ is the dimension of the parameter and $k$ is the
number of machines.
This result holds even in the sparse setting (Braverman et al., 2016).

The method of Jordan, Lee and Yang (2019) proceeds as follows.
Suppose the big data consists of $N$ observations and there are $k$ machines.
For convenience of presentation, we assume that each machine has $n$ observations.
That is, $N=nk$.
Denote the full-data likelihood by
\bas
\ell_N(\theta) = \frac{1}{k} \sum_{j=1}^k \ell_j(\theta),
\eas
where $\ell_j(\theta)$ is the log-likelihood based on the data from the $j$th machine.
For $\theta$ near its target value $\bar \theta$,
\bas
\ell_N(\theta)
&=&
\ell_N(\bar \theta) +
\nabla_{\theta}\ell_N(\theta)\Big|_{\theta = \bar \theta}(\theta - \bar \theta) + R_N(\theta), \\
\ell_1(\theta)
&=&
\ell_1(\bar \theta) +
\nabla_{\theta}\ell_1(\theta)\Big|_{\theta = \bar \theta}(\theta - \bar \theta) + R_1(\theta),
\eas
where $R_N(\theta)$ and $R_1(\theta)$ are remainders.
Observing that  $R_N\approx R_1$, they define a surrogate log-likelihood
\bas
\bar \ell(\theta)
&=&
\ell_N(\bar \theta) +
(\theta - \bar \theta)^\T \nabla_{\theta}\ell_N(\theta)\Big|_{\theta = \bar \theta}
+\left\{ \ell_1(\theta) -
\ell_1(\bar \theta)  -
(\theta - \bar \theta)^\T \nabla_{\theta}\ell_1(\theta)\Big|_{\theta = \bar \theta}\right\}.
\eas
With the constant terms ignored,  the surrogate log-likelihood is
\bas
\bar \ell(\theta)
&=& \ell_1(\theta)   +
\theta^\T\left\{ \nabla_{\theta}\ell_N(\theta)\Big|_{\theta = \bar \theta}
   -
\nabla_{\theta}\ell_1(\theta)\Big|_{\theta = \bar \theta}\right\}.
\eas

The score equation based on the surrogate likelihood is
\[
\nabla_{\theta} \bar \ell (\theta) = \nabla_{\theta} \ell_1(\theta)
+\left\{ \nabla_{\theta}\ell_N(\theta)\Big|_{\theta = \bar \theta}
  -
\nabla_{\theta}\ell_1(\theta)\Big|_{\theta = \bar \theta}\right\}=0.
\]
Let $\hat{\theta}$ be the solution. Expanding it at $\theta_0$ and using the fact that
\[
N^{-1}\{ \nabla_{\theta \theta^\T} \ell_1(\theta_0) - \nabla_{\theta \theta^\T} \ell_N(\theta_0) \}\rightarrow 0  \quad \mbox{in probability.}
\]
Easily we can show that if $\bar\theta-\theta_0=O_p(N^{-1/2})$, then
\[
(\hat{\theta}-\theta_0)=\left\{\nabla_{\theta \theta^\T} \ell_N(\theta_0) \right\}^{-1}
\nabla_{\theta } \ell_N(\theta_0)  +  o_p(N^{-1/2}).
\]
If we let $\bar \theta$ be the MLE based on $\ell_1(\theta)$, then the surrogate log-likelihood can be simplified as
\[
\bar \ell(\theta) =\ell_1(\theta)+ \theta^\T \nabla_{\theta}  \ell_N(\bar \theta),
\]
because $\nabla_{\theta} \ell_1(\bar \theta) =0$.

If the dimension of $\theta$ is high, naturally one may add a penalty function in the surrogate log-likelihood,
and estimate $\theta$ by
\(
\tilde{\theta} = \arg\max_{\theta\in \Theta}\{  \bar{\ell}(\theta) -\lambda||\theta||_1\},
\)
where $||\theta||_1$ is the $L_1$-norm of $\theta$.
Similarly Bayesian inference can be adapted to the surrogate likelihood as well.

Duan et al. (2019) proposed distributed
algorithms which account for the heterogeneous distributions by allowing site-specific nuisance
parameters. The proposed methods extend the surrogate likelihood approach (Wang et al.,
2017; Jordan et al., 2019) to the heterogeneous setting by applying a novel density ratio tilting
method to the efficient score function. It can be shown that asymptotically the approach in Section 6.2 on
nuisance parameters is equivalent to Duan et al. (2019)'s.

\section{Renewal estimation and incremental inference }

Let $U(D_1,\beta) = \nabla_{\beta} M(D_1,\beta)$ be a score function of $\beta$ based on some objective function $M(D_1,\beta)$
from the first batch of data,
where $M$ can either be the log-likelihood
\(
M(D_1,\beta)=\sum_{i=1}^{n_1}\log f(y_{1i}| x_{1i},\beta)
\)
or a log pseudo-likelihood.

Let $\hat{\beta}_1$ be the solution  to
\(
U(D_1,\beta)=0,
\)
when  only the first batch of data  $D_1$ are available.
Let $D_2$ denote the second batch of data.
 If both of them are available, we let
$\hat{\beta}_2$ be the solution to  the pooled score equation,
\(
  U(D_1, \beta)+U(D_2,\beta)=0.
\)
Clearly $\hat{\beta}_2$ is the most efficient estimator of $\beta$
when $D_1$ and $D_2$ are both available.

In addition to $D_2$,
if not $D_1$  but only some summary information $\hat{\beta}_1$ and $\hat{\Sigma}_1$
from it are available,
how to utilize the summary information efficiently?
It is not feasible to estimate $\beta$ by directly solving
\[
U(\beta)  \equiv U(D_1,\beta)+U(D_2,\beta)=0,
\]
which   involves the individual data of the unavailable $D_1$.
Luo and Song (2020) consider  expanding  $U(D_1,\beta)$ at $\beta=\hat{\beta}_1$,  i.e.,
\[
U(D_1,\beta)=U(D_1,\hat{\beta}_1)+   (\beta-\hat{\beta}_1)^\T \nabla_{\beta} U(D_1,\hat{\beta}_1)+
O(\| \beta -\hat{\beta}_1\|^2)
\]
for $\beta$ close to $\hat \beta_1$.
Since $U(D_1,\hat{\beta}_1)=0$, it follows that
\[
U(\beta)= U(D_2,\beta) + (\beta-\hat{\beta}_1)^\T  \nabla_{\beta} U(D_1,\hat{\beta}_1) +
O(\| {\beta}-\hat{\beta}_1\|^2).
\]
Luo and Song (2020) propose to get an updated estimator of $\beta$  by solving
\ba
\label{lu-song}
(\beta-\hat{\beta}_1)^\T  \nabla_{\beta} U(D_1,\hat\beta_1)  +U(D_2,\beta)=0.
\ea

Alternatively we may understand this renewal estimation strategy
in the way of Zhang et al. (2020), who propose to estimate $\beta$ by maximizing
\ba
\label{log-like-9}
\sum_{i=1}^{n_2}\log f(y_{2i}|x_{2i},\beta)-0.5n_1(\hat{\beta}_1-\theta)\Sigma (\hat{\beta}-\beta)^\T
\ea
with
\(
\Sigma= \e \left\{ \nabla_{\beta} \log f(Y|X,\beta) \nabla_{\beta^\T} \log f(Y|X,\beta) \right\}
\)
being
 the Fisher information.
If both batches are available, the score for $\beta$ is
\[
S(\beta)=\sum_{i=1}^{n_1} \nabla_{\beta} \log f(y_{1i}|x_{1i},\beta) +  \sum_{i=1}^{n_2} \nabla_{\beta}   \log f(y_{2i}|x_{2i},\beta).
\]
After recording  $\hat{\beta}_1$ and $\Sigma$, we do not have the
raw data $ \{(y_{1i},x_{1i}), i=1,2,...,n_1\}$ anymore.
Because
\begin{eqnarray*}
\hat{\beta}_1-\beta
= -n_1^{-1}\Sigma^{-1} \sum_{i=1}^{n_1} \nabla_{\beta}  \log f(y_{1i}|x_{1i},\beta) +o_p(n_1^{-1/2}),
\end{eqnarray*}
differentiating \eqref{log-like-9}
with respect to $\beta$ gives
\begin{eqnarray*}
&&\hspace{-2cm}\sum_{i=1}^{n_2} \nabla_{\beta} \log f(y_{2i}|x_{2i},\beta) -n_1\Sigma(\hat{\beta}_1-\beta)\\
&=& \sum_{i=1}^{n_1} \nabla_{\beta} \log f(y_{1i}|x_{1i},\beta) +\sum_{i=1}^{n_2} \nabla_{\beta}\log f(y_{2i}|x_{2i},\beta) +o_p(n^{1/2}).
\end{eqnarray*}
Here we have assumed that $n_1=O(n_2)=O(n)$.
This indicates that  estimating $\beta$ by
maximizing  \eqref{log-like-9} has no efficiency loss
asymptotically  compared with  the MLE based on   all individual data,
where the latter is infeasible in the current situation.

\section{Concluding remarks}

{Rapid growth in hardware technology has made the data collection much easier and more effectively.
In many applications,  data often arrive in streams and chunks,
which leads to batch by batch data or streaming data.
For example,   web sites severed
by widely distributed web servers may need to coordinate
many distributed clickstream analyses, e.g. to track
heavily accessed web pages as part of their real-time performance
monitoring.
Other examples  include financial applications,
network monitoring, security, telecommunications data management,
  manufacturing, and sensor networks (Babcock et al., 2002; Nguyen et al., 2020).
The  continuous arrival of such data in multiple, rapid, time-varying,
possibly unpredictable and unbounded streams   yields some
fundamentally new research problems.
One of the most challenging issues is   how
to address statistics in an online updating framework, without storage requirement
for  raw data.
}

Assembling information from difference data sources
has become indispensable in big data and artificial
intelligence research.
Statistical tools play an essential role in updating information.
In this paper,  we have made a selective review on
several traditional statistical methods,
such as  meta analysis,  calibration information methods in survey sampling,
EL together with over-identified  estimating equations,
and GMM.  We also briefly review some recently-developed statistical methods,
including   communication efficient distributed statistical inference
 and renewal estimation and incremental inference,
which can be regarded as the latest development of calibration information methods
in the era of big data. Even though those methods were developed in different fields and in different statistical frameworks,
in principle, they are asymptotically equivalent to those well known methods developed in meta analysis.
Almost no or little information is lost compared with the case when full data are available.

Due to deficiency of our knowledge, finally we have to apology for individuals whose works inadvertently have been left off in our reference lists.

\section*{References}
\begin{description}

\item
Babcock, B., Babu, S., Datar, M., Motwani, R., and Widom, J. (2002).
Models and issues in data stream systems. In: {\it
 Proceedings of the twenty-first ACM SIGMOD-SIGACT-SIGART
 symposium on Principles of database systems}, 1--16.

\item
Back, K. and Brown, D. P. (1992). GMM, maximum likelihood,
and nonparametric efficiency. {\it Economics Letters}, {\bf 39}, 23--28.

\item
Braverman, M., Garg, A., Ma, T., Nguyen, H., and Woodruff, D. (2016).
Communication lower bounds for statistical estimation problems
via a distributed data processing inequality.  In: {\it Proceedings of the
Forty-Eighth Annual ACM Symposium on Theory of Computing}, New
York: ACM, 1011--1020.

\item
Borenstein, M.,  Hedges, L. V.,  Higgins, J. P. T., and  Rothstein, H. (2009).
{\it Introduction to Meta-Analysis}. Chichester: Wiley.

\item
Chatterjee, N., Chen, Y.-H., Maas, P., and Carroll, R. J. (2016).
Constrained maximum likelihood estimation for model calibration
using summary-level information from external big data sources.
{\it Journal of the American Statistical Association}, {\bf 111}, 107--117.

\item
Chaudhuri, S., Handcock, M. S., and Rendall, M. S. (2008).
Generalized linear models incorporating population level information:
an empirical likelihood based approach.
{\it Journal of the Royal Statistical Society: Series B}, {\bf 70}, 311--328.

\item
Chen, J. and  Qin, J. (1993).
Empirical likelihood estimation for finite populations and the effective
usage of auxiliary information.
{\it Biometrika}, {\bf 80}, 107--116.

\item
Chen, J., Sitter, R., and Wu, C. (2002).
Using empirical likelihood methods to obtain range restricted
weights in regression estimators for surveys. {\it Biometrika}, {\bf 89}, 230--237.

\item
Cochran, W. G. (1977). {\it Sampling Techniques}. 3rd Edition. New York: Wiley.

\item
Dersimonian, R.  and Laird, N.  (1986).
Meta-analysis in clinical trials.
{\it Controlled Clinical Trials}, {\bf 7}, 177--188.

\item
Duan, R.,  Ning, Y., and Chen, Y. (2020).
Heterogeneity-aware and communication-efficient distributed statistical inference.
{\it ArXiv:1912.09623v1}.

\item
Duchi, J., Jordan, M., Wainwright, M., and Zhang, Y. (2015).
Optimality guarantees for distributed statistical estimation.
{\it arXiv:1405.0782}.

\item
Han, P and Lawless, J. (2016)
Comment.
{\it Journal of the American Statistical Association}, {\bf 111},  118--121.

\item
 Hansen, L. P.  (1982).
 Large sample properties of generalized method of moments estimators.
{\it Econometrica}, {\bf 50}: 1029--1054.

\item
Hartely, H. O. and Rao, J. N. K. (1968). A new estimation theory for sample
surveys. {\it Biometrika }, {\bf 55}, 547--557.

\item
Huang, C. Y. and Qin, J. (2020).
A unified approach for synthesizing population level covariate
effect information in semiparametric estimation with survival data.
{\it Statistics in Medicine}. In press.

\item
Imbens, G. and Lancaster, T (1994).
Combining micro and macro data in microeconometric models.
{\it Review of Economic Studies}, {\bf 61}, 655--680.

\item
Jordan, M. I., Lee, J. D.,  and Yang, Y. (2019).
 Communication-efficient distribution statistical inference.
{\it Journal of the American Statistical Association}, {\bf 114}, 668--681.

\item
Lee, J., Liu, Q., Sun, Y.,  and Taylor, J. (2017).
  Communication-efficient sparse regression.
  {\it Journal of Machine Learning Research}, {\bf 18},  1--30.

\item
Lin, D. Y. and Zeng, D. (2010).
On the relative efficiency of using summary statistics versus
individual-level data in meta-analysis.
{\it Biometrika},  {\bf 97}, 321--332.

\item
Luo, L. and Song, P. X. K. (2020). Renewable estimation and incremental inference in generalized linear models with streaming data sets.
{\it Journal of the Royal Statistical Society, Series B}, {\bf 82}, 69--97.

\item
Neiswanger, W., Wang, C., and Xing, E. (2015).
Asymptotically exact, embarrassingly parallel MCMC.
In: {\it Proceedings of the Thirtieth Conference
on Uncertainty in Artificial Intelligence}, Arlington, VA: AUAI
Press, 623--632.

\item
Nguyen, T. D.,  Shih, M. H.,  Srivastava, D.,   Tirthapura, S., and Xu, B. (2020).
Stratified random sampling from streaming and stored
data. {\it  Distributed and Parallel Databases}.
In press.

\item
Owen, A. B. (1988).
Empirical likelihood ratio confidence intervals for a single functional.
{\it Biometrika}, {\bf 75}, 237--249.

\item
Owen, A. B. (1990).
Empirical likelihood ratio confidence regions.
{\it The Annals of Statistics}, {\bf 18},  90--120.

\item
Owen, A. B. (2001). {\it Empirical Likelihood}. New York: CRC.

\item
Qin, J. (2000). Combining parametric and empirical likelihoods.
{\it Biometrika}, {\bf 87}, 484--490.

\item Qin, J. (2017). {\it Biased sampling, over-identified parameter problems and beyond.} Singapore: Springer-Verlag.

\item
Qin, J. and Lawless, J. (1994).
Empirical likelihood and general equations.
{\it The Annals of Statistics}, {\bf 22}, 300--325.

\item
Qin, J., Zhang, H., Li, P., Albanes, D., and Yu, K. (2015).
Using covariate specific disease prevalence information to
increase the power of case-control study.   {\it Biometrika}, {\bf 102}, 169--180.

\item
Susanne,  M. S. (2007).
Point estimation with exponentially tilted empirical likelihood.
{\it  The Annals of Statistics},  {\bf 35},  634--672.

\item
Tian, L. and Gu Q. (2016). Communication-efficient distributed sparse linear discriminant analysis.
{\it arXiv:1610.04798}.

\item
van de Geer, S., Buhlmann, P., Ritov, Y., and Dezeure, R. (2014).
On asymptotically optimal confidence regions and tests for high dimensional
models.{\it The Annals of Statistics}, {\bf 42}, 1166--1202

\item
Wang, X., and Dunson, D. (2015).
Parallelizing MCMC via Weierstrass Sampler.
{\it arXiv:1312.4605}.

\item
Wang, J., Kolar,  M.,  Srebro, N.,  and  Zhang, T. (2017). Efficient distributed learning with sparsity.
In: {\it  Proceedings of the 34th International Conference on Machine Learning, PMLR 70}, 3636--3645.

\item
Wang, X., Yang, Z. ,  Chen, X.,  and  Liu, W. (2019).
Distributed inference for linear support vector machine.
{\it Journal of Machine Learning Research},  {\bf 20}, 1--41.

\item
Wu, C. and Sitter, R. R. (2001). A model-calibration approach to using complete auxiliary information from survey data.
{\it Journal of the American Statistical Association}, {\bf 96}, 185-193.

\item
Wu, C. and Thompson, M. E. (2020).
{\it Sampling Theory and Practice}.
Switzerland: Springer.

\item
Zhang, Y., Duchi, J., and Wainwright, M. (2013).
Communication-efficient algorithms for statistical optimization.
{\it Journal of Machine Learning Research}, {\bf 14}, 3321--3363.

 \item
  Zhang, H., Deng, L., Schiffman, M., Qin, J.,  and  Yu, K. (2020).
 Generalized integration model for improved statistical inference by leveraging external summary data.
 {\it Biometrika}, {\bf 107}, 689-703.

\item
Zhao, T.,  Cheng, G.,   and  Liu, H. (2016).
A partially linear framework for massive heterogeneous data.
{\it The Annals of Statistics}, {\bf 44}, 1400--1437.
\end{description}

\end{document}